\documentclass[a4paper,USenglish,cleveref, autoref, thm-restate]{lipics-v2021}

\pdfoutput=1 
\hideLIPIcs  

\title{Congestion bounds via Laplacian eigenvalues and their application to tensor networks with arbitrary geometry}

\titlerunning{Congestion and Laplacian eigenvalues} 

\author{Sayan Mukherjee}{Department of Physics, The University of Tokyo, Bunkyo-ku, Tokyo, 113-0033, Japan \and \url{http://sayan.mukherjee.moe} }{sayan@phys.s.u-tokyo.ac.jp}{https://orcid.org/0000-0001-8838-0455}{Research supported by JSPS KAKENHI Grant Number 24K22830, and the Center of Innovations for Sustainable Quantum AI (JST Grant Number JPMJPF2221).}

\author{Shinichiro Akiyama}{Center for Computational Sciences, University of Tsukuba, Tsukuba, Ibaraki 305-8577, Japan \and Department of Physics, The University of Tokyo, Bunkyo-ku, Tokyo, 113-0033, Japan \and \url{https://akiyama-es.github.io}}{akiyama@ccs.tsukuba.ac.jp}{https://orcid.org/0000-0003-1415-5620}{Research supported by JSPS KAKENHI Grant Numbers 23K13096 and 25H01510, the Center of Innovations for Sustainable Quantum AI (JST Grant Number JPMJPF2221), and the Top Runners in Strategy of Transborder Advanced Researches (TRiSTAR) program conducted as the Strategic Professional Development Program for Young Researchers by the MEXT.}

\authorrunning{S. Mukherjee and S. Akiyama} 

\Copyright{Sayan Mukherjee and Shinichiro Akiyama} 



\category{} 

\relatedversion{} 

\nolinenumbers 

\newcommand{\overbar}[1]{\mkern 1.5mu\overline{\mkern-1.5mu#1\mkern-1.5mu}\mkern 1.5mu}
\newcommand{\ContractionTree}{\mathcal T}

\newcommand{\tw}{\mathop{\mathrm{tw}}}
\newcommand{\cng}{{\mathrm{cng}}}
\newcommand{\Vol}{{\mathop{\mathrm{Vol}}}}

\newcommand{\argmin}{\mathop{\mathrm{argmin}}}

\usepackage[linesnumbered,ruled,vlined]{algorithm2e}
\usepackage{amsfonts, amsmath, amssymb, amsthm, tikz}

\theoremstyle{definition}
\newtheorem{cor}{Corollary}

\newtheorem{prop}{Proposition}
\newtheorem{defn}{Definition}
\newtheorem{alg}[theorem]{Algorithm}

\EventEditors{John Q. Open and Joan R. Access}
\EventNoEds{2}
\EventLongTitle{42nd Conference on Very Important Topics (CVIT 2016)}
\EventShortTitle{CVIT 2016}
\EventAcronym{CVIT}
\EventYear{2016}
\EventDate{December 24--27, 2016}
\EventLocation{Little Whinging, United Kingdom}
\EventLogo{}
\SeriesVolume{42}
\ArticleNo{23}

\begin{document}

\maketitle

\begin{abstract}
    Embedding the vertices of arbitrary graphs into trees while minimizing some measure of overlap is an important problem with applications in computer science and physics.
    In this work, we consider the problem of bijectively embedding the vertices of an $n$-vertex graph $G$ into the \emph{leaves} of an $n$-leaf \emph{rooted binary tree} $\ContractionTree$.
    The congestion of such an embedding is given by the largest size of the cut induced by the two components obtained by deleting any vertex of $\ContractionTree$.
    We show that for any embedding, the congestion lies between $\lambda_2(G)\cdot 2n/9$ and $\lambda_n(G)\cdot n/4$, letting $0=\lambda_1(G)\le \cdots \le \lambda_n(G)$ be the Laplacian eigenvalues of $G$, and there is an embedding for which the congestion is at most $\lambda_n(G)\cdot 2n/9$.
    Beyond these general bounds, we determine the congestion exactly for hypercubes and lattice graphs, and obtain asymptotically tight bounds for random regular graphs and Erd\H os–R\'enyi graphs.
    We further introduce an efficient contraction procedure based on spectral ordering and dynamic programming, which produces low-congestion embeddings in practice. Numerical experiments on structured graphs, random graphs, and tensor network representations of quantum circuits validate our theoretical bounds and demonstrate the effectiveness of the proposed method.
    These results yield new spectral bounds on the memory and time complexity of exact tensor network contraction in terms of the underlying graph structure.

\end{abstract}


\section{Introduction}
\label{sec:introduction}
In recent years, tensor networks have become an irreplaceable tool in various areas of science.
Numerous applications have been found in fields such as quantum many-body physics~\cite{PhysRevLett.69.2863,Verstraete_2008,Orus:2018dya,Okunishi:2021but}, quantum circuit simulation~\cite{PhysRevLett.91.147902,Huang:2020ptj,PhysRevResearch.3.023005,10.1145/3458817.3487399}, and machine learning~\cite{stoudenmire2016supervised,PhysRevB.97.085104,PhysRevB.99.155131,Martyn:2020kzq}.
The basic structure of such applications involves representing the desired quantity through the \textit{contraction} of a network of tensors.
Tensor networks with regular or repeating structures, such as lattices, can usually be efficiently computed using so-called renormalization group algorithms~\cite{PhysRevLett.69.2863,Nishino_1996,Levin:2006jai,Meurice:2020pxc,Okunishi:2021but}. At the same time, analyzing the complexity of contracting tensor networks with arbitrary geometries remains an important avenue of research for extending their applicability, as it is crucial for developing efficient computational algorithms for inference and learning problems on general graphs, as well as for classical simulations of quantum circuits with arbitrary connectivity~\cite{PhysRevX.14.011009,10.21468/SciPostPhys.8.1.005,PhysRevLett.125.060503}.
{Additionally, statistical models on random graphs are closely related to the study of quantum gravity~\cite{Eguchi:1981kk,Ambjorn:1985az,Knizhnik:1988ak,Kazakov:1995ae} and random tensor networks offer a novel approach to realizing holographic duality also in the context of high-energy physics~\cite{Hayden:2016cfa}.}

In its mathematical form, a tensor network $(G, T)$ can be represented by a weighted graph $G$ with vertex set $V(G)$ and edge set $E(G)$, and tensors $T=\{T_v: v\in V(G)\}$ on its vertices, where an edge of weight $\log_2\chi$ between nodes $u$ and $v$ represents a common index between $T_u$ and $T_v$ of dimension $\chi$.
Given two nodes of a tensor network $u$ and $v$, the contraction is equivalent to summing over all the common edges between $u$ and $v$.
The order in which nodes of the tensor network are contracted can give rise to huge variations in the memory and time complexity.
Therefore, finding a contraction order that minimizes this complexity is an important avenue of research from both practical and computational theory perspectives.

As is the case for several important problems in computational science, the question of finding a good contraction procedure is known to be \#P-hard~\cite{TNContractionSharpP-damm-holzer-mckenzie-2002}, that is, it is at least hard as counting the number of solutions to any NP-hard decision problem.
Regardless, procedures such as dynamic programming on small graphs~\cite{optEinsum-pfeifer-haegeman-verstraete-2014}, tree decomposition based procedures~\cite{SimulatingQuantumComp-markov-shi-2008,HeuristicsTreewidthTrimming-Schutski-Khakulin-2020}, greedy or other graph partitioning procedures~\cite{HyperOptimizedTNContraction-gray-kourtis-2021}, decision diagrams~\cite{TNDecisionDiagram-Burgholzer-Ploier-Wille-2023,TNDecisionDiagram-Hong-Zhou-Li-Feng-Ying-2022}, planar graph separators~\cite{FastCounting-Kourtis-Chamon-Mucciolo-Ruckenstein-2019}, genetic algorithms~\cite{GeneticAlgorithmTNC-Frank-Adam-2020} and so on are used in practice.
Sub-exponential upper bounds have been shown in the theoretical studies of \cite{FastCounting-Kourtis-Chamon-Mucciolo-Ruckenstein-2019} and \cite{sphereSeparatorTNC-Wahl-Sergii-2023}.
See also \cite{ParametrizationTNContracition-d-gorman-2019} for an analysis of parameterizing the cost of tensor network contraction.

In this work, we focus on the memory requirement of contraction of general tensor networks from the perspective of graph theory.
A natural way of representing a contraction order of a tensor network $(G, T)$ is via a \textit{rooted binary tree} $\ContractionTree$ together with a bijection $\pi$ from the nodes of $G$ to the leaves of $\ContractionTree$ (see \cite{ParametrizationTNContracition-d-gorman-2019,HyperOptimizedTNContraction-gray-kourtis-2021,sphereSeparatorTNC-Wahl-Sergii-2023}).
Let $e_G(X,Y)$ denote the number of edges between $X\subseteq V(G)$ and $Y\subseteq V(G)$.
Then every node $B$ of $\ContractionTree$ corresponds to a subset $\pi^{-1}(B)\subseteq V(G)$, and is an intermediate tensor of rank given by the cut size $e_G(\pi^{-1}(B), \overbar{\pi^{-1}(B)})$.
The largest rank of any intermediate tensor dominates the memory cost incurred during the contraction procedure, and is known as the edge-congestion of the contraction order $(\pi, \ContractionTree)$.
We illustrate an example tensor network and its contraction order representation in Fig.~\ref{fig:intro-contraction-as-binary-tree}.

\begin{figure}[h]
\centering
    \begin{tikzpicture}
        \tikzstyle{vertex}=[align=center, inner sep=0pt, text centered, circle,black,fill=gray!10,draw=black,text width=1.5em,thick]
        \tikzstyle{treevertex}=[rectangle,fill=gray!10,draw=black,minimum size=2pt,inner sep=1.3pt]
        \node [vertex] (x1) at (0:2){$1$};
        \node [vertex] (x2) at (60:2){$2$};
        \node [vertex] (x3) at (120:2){$3$};
        \node [vertex] (x4) at (180:2){$4$};
        \node [vertex] (x5) at (240:2){$5$};
        \node [vertex] (x6) at (300:2){$6$};
        \draw (x1)--(x2)--(x3)--(x6)--(x4)--(x3)--(x5)--(x6);
        \begin{scope}[shift={(7, -0.5)}]
            \node [treevertex] (v) at (0,2){$\{1,2,3,4,5,6\}$};
            \draw (v) node [above=0.5em] {\scriptsize{$0$}};
            \node [treevertex] (v1) at (-2,1){$\{1,2\}$};
            \node [treevertex] (v2) at (2,1){$\{3,4,5,6\}$};
            \draw (v1) node [above=0.5em] {\scriptsize{$1$}};
            \draw (v2) node [above=0.5em] {\scriptsize{$1$}};
            \node [treevertex] (x1) at (-2.5, 0){$\{1\}$};
            \node [treevertex] (x2) at (-1.5, 0){$\{2\}$};
            \draw (x1) node [above=0.5em] {\scriptsize{$1$}};
            \draw (x2) node [above=0.5em] {\scriptsize{$2$}};
            \node [treevertex] (v3) at (0.5, 0){$\{3,4\}$};
            \node [treevertex] (v4) at (3.5, 0){$\{5,6\}$};
            \draw (v3) node [above=0.5em] {\scriptsize{$4$}};
            \draw (v4) node [above=0.5em] {\scriptsize{$3$}};
            \node [treevertex] (x3) at (0,-1){$\{3\}$};
            \node [treevertex] (x4) at (1,-1){$\{4\}$};
            \node [treevertex] (x5) at (3,-1){$\{5\}$};
            \node [treevertex] (x6) at (4,-1){$\{6\}$};
            \draw (x3) node [above=0.5em] {\scriptsize{$4$}};
            \draw (x4) node [above=0.5em] {\scriptsize{$2$}};
            \draw (x5) node [above=0.5em] {\scriptsize{$2$}};
            \draw (x6) node [above=0.5em] {\scriptsize{$3$}};
            \draw (x1)--(v1)--(x2);
            \draw (x3)--(v3)--(x4);
            \draw (x5)--(v4)--(x6);
            \draw (v1)--(v)--(v2);
            \draw (v3)--(v2)--(v4);
        \end{scope}
    \end{tikzpicture}
    \caption{\label{fig:intro-contraction-as-binary-tree}(Left): A tensor network $(G,T)$ on $6$ nodes where each bond of $T$ has dimension $2$. (Right): The rooted binary tree $\mathcal B$ representing a contraction order. A node $S$ of $\mathcal B$ represents an intermediate tensor encountered during the contraction procedure and has rank $|\partial S|$, which is indicated at the top of each node. The edge-congestion of this embedding is $4$.} 
\end{figure}

To the knowledge of the authors, the first comprehensive study of this particular embedding was that of Bienstock in 1990~\cite{embeddingGraphsInTrees-bienstock-1990}.
They defined two notions of congestion by ``routing'' edges $uv\in E(G)$ via paths through the binary tree $\ContractionTree$.
The largest number of paths sharing a common edge of $\ContractionTree$ is known as the edge-congestion (which is equivalent to the definition above!), and sharing a common vertex of $\ContractionTree$ the vertex-congestion.
Gorman~\cite{ParametrizationTNContracition-d-gorman-2019} showed that edge-congestion determines the memory complexity, and vertex-congestion determines the time complexity of tensor network contraction.
However, it was also known (Remark 3 of the Appendix of \cite{embeddingGraphsInTrees-bienstock-1990}) that the vertex-congestion is within $1$ and $1.5$ times the edge-congestion, implying that bounds on edge-congestion naturally lead to (within constant-factor) bounds on vertex-congestion.
In this work, we focus on how spectral properties of $G$ affect the edge-congestion, and by $\cng(G)$ or \emph{congestion} of $G$, simply mean the edge-congestion of $G$.

We now formalize our notation, following \cite{embeddingGraphsInTrees-bienstock-1990}.
Given a weighted graph $G$ representing a tensor network, and a contraction order $(\pi, \ContractionTree)$, the congestion of $G$ under $(\pi,\ContractionTree)$ is defined as $\cng_{\pi, \ContractionTree}(G) = \max_{B\in V(\ContractionTree)}\, |\partial_G(\pi^{-1}(B))|$.
Define the congestion of $G$ as,
\begin{equation}
\label{eq:intro-congestion-def}
    \cng(G):=\min_{(\pi,\ContractionTree)}\, \max_{B\in V(\ContractionTree)}\, \cng_{\pi, \ContractionTree}(B).
\end{equation}

The problem of finding low-congestion embeddings of $G$ has recently resurfaced in the work of Harvey and Wood~\cite{TreewidthLineGraph-harvey-wood-2018}.
In particular, they relate the congestion of a graph with the treewidth of its line graph
\footnote{The line graph of $G$, denoted by $L(G)$, has vertex set $E(G)$, and $e_1$ and $e_2$ are adjacent in $L(G)$ iff they share an endpoint. The treewidth of a graph is the size of the largest vertex set in a tree decomposition~\cite{TreewidthMinorsII-robertson-seymour-1986}.}
$\tw(L(G))$ via:
\begin{equation}
\label{eq:intro-congestion-tw(L(G))}
    \cng(G)=\tw(L(G))+1,
\end{equation}
and prove several bounds on $\tw(L(G))$ depending on the degrees of the nodes in $G$.
It is worth noting that (\ref{eq:intro-congestion-tw(L(G))}) was already known in the physics community as a restatement of the work of Markov and Shi~\cite{SimulatingQuantumComp-markov-shi-2008}, where they obtain a low-rank tensor network contraction procedure of $G$ from tree decompositions of $L(G)$.

For a multigraph $G$ without self-loops, the Laplacian matrix is defined as $L_G=D_G-A_G$, where $D_G$ is the diagonal degree matrix and $A_G$ is the adjacency matrix.
It is known that $L_G$ is positive semi-definite, has eigenvalues $0=\lambda_1(G)\le \cdots\le \lambda_n(G)\le 2\Delta(G)$, where $\Delta(G)$ is the maximum degree of any node in $G$.
The spectrum of $L_G$ plays an important role in determining the connectivity and clustering properties of $G$, and forms the basis of the highly practical spectral clustering algorithm~\cite{Isoperimetric-mohar-1989,SpectralClustering-ng-jordan-weiss-2001}.
For more information on the graph Laplacian, the reader is referred to the survey article \cite{LaplacianSurvey-mohar-1991}.

\subsection*{Our Results}


In this work, we establish lower and upper bounds on congestion of arbitrary graphs in terms of their Laplacian spectra, determine its exact or asymptotic value for several fundamental graph families, and derive algorithmic consequences for tensor network contraction.
To describe our general bound, we first define a quantity $\varepsilon(G)$, which can be interpreted as the balance of the spectral clustering output on $G$.
\begin{defn}[Spectral balance $\varepsilon(G)$]
\label{defn:VarEpsilon}
For a graph $G$, let $x$ be the eigenvector of $\lambda_2(G)$ of $L_G$.
Suppose $V(G)=S_+\sqcup S_-$ is an equitable partition of $V(G)$ with $x_i\ge0$ for all $i\in S_+$, and $x_i\le 0$ for all $i\in S_-$.
Then,
\begin{equation}
\label{eq:intro-varepsilon-definition}
    \varepsilon(G):=\frac{\min\{|S_+|, |S_-|\}}{n}.
\end{equation}
\end{defn}

\begin{theorem}[Main Theorem]
    \label{thm:MAINTHM}
    For any graph $G$ and embedding $\pi:V(G)\to\ell(\ContractionTree)$, the following holds.
    \begin{equation}
        \label{eq:intro-thm-mainthm}
        \frac{2\lambda_2(G)}9\le \frac{\cng_{\pi, \ContractionTree}(G)}n\le \frac{\lambda_n(G)}4.
    \end{equation}
    Moreover, there are two embeddings $\pi_i:V(G)\to\ell(\ContractionTree_i)$, $i=1,2$, such that
    \begin{align}
        \frac{\cng_{\pi_1, \ContractionTree_1}(G)}n&\le \frac{2\lambda_n(G)}{9}, \text{ and } \label{eq:intro-thm-mainthm-3}\\
        \frac{\cng_{\pi_2, \ContractionTree_2}(G)}n&\le \max\left\{\varepsilon(G) \sqrt{(2\Delta(G)-\lambda_2(G))\lambda_2(G)}\, ,\,  \left(\frac{1-\varepsilon(G)^2+1/n}4\right) \lambda_n(G)\right\}. \label{eq:intro-thm-mainthm-2}
    \end{align}
    
\end{theorem}

Although of theoretical interest and \emph{asymptotically} tight in many situations, in practice, the upper bounds of (\ref{eq:intro-thm-mainthm-3}) and (\ref{eq:intro-thm-mainthm-2}) are very loose.
Therefore, we propose a simple contraction heuristic which typically achieves low congestion, and in some cases performs better than competing methods such as the Cotengra-Auto and Hyper-Greedy heuristics of Gray and Kourtis~\cite{HyperOptimizedTNContraction-gray-kourtis-2021}.
This method orders the vertices of $G$ by spectral ordering and then applies the dynamic programming approach of Ref.~\cite{OptimalContractionTrees-ibrahim-et-al-2022} to find a lowest congestion contraction tree.
Our algorithm has complexity $O(n^3)$.

\begin{alg}[Spectral order and dynamic programming contraction tree]
\label{def:Algo-SO-DP}
Let $G$ be a graph and let $x$ be an eigenvector of $\lambda_2(G)$.
Let $\sigma$ be the permutation of $V(G)$ such that
$x_{\sigma(1)}\le \cdots \le x_{\sigma(n)}$ ($\sigma$ is known as the \emph{spectral ordering} of $G$).
For $1\le i\le j\le n$, write $S_{i,j} := \{\sigma(i),\ldots,\sigma(j)\}$.
Define the \emph{spectral DP cost} $C(i,j)$ recursively by $C(i,i):= \deg_G(\sigma(i))$, and for $1\le i<j\le n$,
\[
C(i,j) := \min_{i\le k<j} \max\Bigl(
  C(i,k),\,
  C(k+1,j),\,
  e_G\bigl(S_{i,k},V(G)\setminus S_{i,k}\bigr),\,
  e_G\bigl(S_{k+1,j},V(G)\setminus S_{k+1,j}\bigr)
\Bigr),
\]

The \emph{spectral DP congestion} of $G$ is $C(1,n)$.
A corresponding contraction tree is obtained by recording, for each interval $[i,j]$, an index $k^\ast(i,j)$ attaining the minimum in the definition of $C(i,j)$, and recursively splitting $S_{i,j}$ into $S_{i,k^\ast}$ and $S_{k^\ast+1,j}$.

Since the second eigenvector approximately reveals balanced sparse cuts, vertices likely to be separated by small cuts are placed close together in the spectral ordering, motivating the choice of $\sigma$.
\hfill{$\blacksquare$}
\end{alg}

We also analyze the congestion of several graph families used in computer science and physics: hypercubes, lattices, and random graphs.
We show that the congestion of the $d$-dimensional hypercube is $2^{d-1}$ (Proposition~\ref{prop:hypercube-exact-cng}) and that the congestion of the periodic and non-periodic $m\times n$ rectangular lattices are given by $2\min\{m,n\}$ and $\min\{m,n\}$, respectively (Proposition~\ref{prop:lattice-exact-cng}).
Finally, we prove asymptotically tight bounds on congestion of random regular graphs $\mathcal G(n,d)$ and Erd\H os-R\'enyi random graphs $\mathcal G_{n,p}$.
\begin{theorem}
\label{thm:RRG-ERG-Congestion}
    For large $n$, with probability $1-o(1)$, it holds that $\cng(G)=\frac 29nd+O(\sqrt{nd\log n})$ for $G\sim \mathcal G(n,d)$ and $\cng(G)=\frac 29pn^2 + O(\sqrt{pn^2\log n})$ for $G\sim \mathcal G_{n,p}$.
\end{theorem}

\subsection*{Corollaries}

Theorem~\ref{thm:MAINTHM} directly implies the following corollaries on the \emph{exact} contraction complexity for general tensor networks, and on treewidth.
\begin{cor}
    \label{cor:intro-TN-cor}
    Given any tensor network on an $n$-vertex multigraph $G$ without self loops, and under no assumptions about the individual tensors, any contraction order will always create a tensor of rank at least $2\lambda_2(G)n/9$.
    On the other hand, one can always find a contraction order such that the largest rank tensor produced by the order is at most $\min\,\{\mathrm{RHS}(\ref{eq:intro-thm-mainthm-3}), \mathrm{RHS}(\ref{eq:intro-thm-mainthm-2})\}$.
\end{cor}

Although Corollary~\ref{cor:intro-TN-cor} applies to exact contraction, the spectral lower bound identifies unavoidable high-rank intermediate tensors and thus helps guide approximate or sliced contraction schemes.

\begin{cor}
    \label{cor:intro-treewidth-lambda2-cor}
    Another corollary of the upper bound in (\ref{eq:intro-thm-mainthm}) together with (\ref{eq:intro-congestion-tw(L(G))}) and the fact that $\tw(G)\le \cng(G)$ (see (\ref{eq:intro-tw-L(G)-Delta(G)})), is an upper bound of $\min\,\{\mathrm{RHS}(\ref{eq:intro-thm-mainthm-3}), \mathrm{RHS}(\ref{eq:intro-thm-mainthm-2})\}$ on $\tw(G)$ in terms of the Laplacian spectrum of $G$.
\end{cor}

To the best of our knowledge, while there has been previous work on finding lower bounds on $\tw(G)$ depending on Laplacian eigenvalues \cite{spectralTreewidthLowerBd-chandran-subramanian-2003, SpectralTWLB-gima-hanaka-noro-ono-otachi-2024, SpectralTWLB-gima-hanaka-noro-ono-otachi-2024-arXiv}, we provide the first theoretical improvement from the trivial bound of $\lambda_n(G)\cdot n/4$.
Note that in practice, efficient algorithms computing tree decompositions are abundant; however, Corollary~\ref{cor:intro-treewidth-lambda2-cor} is mainly of theoretical interest.
Moreover, most practical tensor network applications focus on \emph{approximate} contraction heuristics, whereas the goal of this work is to explore theoretical limits of \emph{exact} contraction algorithms.

\subsection*{Organization}
This paper is organized as follows.
Sec.~\ref{sec:background-related-work} covers background and related works.
Sec.~\ref{sec:congestion-bounds-via-laplacian} proves Theorem~\ref{thm:MAINTHM}, and provides an alternative version using the normalized Laplacian matrix,
which is better in situations with high degree variations in $G$.
We determine $\cng(G)$ for several graph families used in computer science and physics in Sec.~\ref{sec:theoretical-congestion-bounds}.
Finally, we numerically compare the performance of Algorithm~\ref{def:Algo-SO-DP} and the contraction procedures of Gray and Kourtis~\cite{HyperOptimizedTNContraction-gray-kourtis-2021} and Ibrahim et. al.~\cite{OptimalContractionTrees-ibrahim-et-al-2022} in Sec.~\ref{sec:experiments}.
Sec.~\ref{sec:conclusion-future-work} is devoted to the conclusion and outlook.

\section{Background and Related works}
\label{sec:background-related-work}
\subsection{Graph embedding}
Embedding general graphs into trees (and more generally, other graphs) with small congestion is an important problem with many practical applications in routing and bandwidth problems~\cite{bandwidthProblemSurvey-chinn-chatalov-dewdney-gibbs-1982,treewidthPathwidthCutwidth-korach-solel-1993}, design and visualization of Very Large Scale Integrated Chips (VLSI)~\cite{VLSIFramework-bhatt-leighton-1984,planarDAGStackNumber-nollenberg-pupyrev-2023,maximumSubgraphSierpinski-rajan-greeni-joswa-2023}, parallel algorithms design~\cite{embeddingMeshes-barth-1993,leighton2014introduction}, tensor network contraction~\cite{HyperOptimizedTNContraction-gray-kourtis-2021,SimulatingQuantumComp-markov-shi-2008}, to cite only a handful from the vast amount of applications in physics and computer science.

Similar concepts of congestion have been studied in many settings, including decision problems for tree embeddings, embeddings into spanning trees, and embeddings of specific graph families such as hypercubes~\cite{optimalArrangementDataTree-luczak-noble-2001,minimalCongTrees-ostrovskii-2004,MinimumAvgCong-manuel-2011,spanningTreeCongestion-kozawa-otachi-yamazaki-2009,spanningTreeCongestion-lowenstein-rautenbach-regen-2009}.

\subsection{Carving width}
When $\ContractionTree$ is an \emph{unrooted binary tree}, i.e. all nodes have degree $3$ or $1$, an analogous width parameter known as carving width was introduced by Seymour and Thomas~\cite{Ratcatcher-seymour-1994}.
In this case, for any embedding $\pi:G\to\ell(\ContractionTree)$, deleting an edge of $\ContractionTree$ induces a partition of $V(G)$, and the carving width of $G$ with respect to $(\pi, \ContractionTree)$ is defined as the largest cut size of any such partition.
By definition, $\cng(G)$ is always at least the carving width for any $G$.
The effect of carving width in representing the complexity of tensor network contraction has been studied in \cite{carvingWidthTN-jakes-2019}, and in \cite{QFFTClassicalSimulated-aharonov-2006} under the name of ``bubble width.''

\subsection{Treewidth and Laplacian eigenvalues}
\label{sec:treewidthLaplacianPreviousWork}
\paragraph*{Treewidth of the original graph.}
There is a line of investigation closely related to ours in studying $\tw(G)$ with the algebraic connectivity $\lambda_2(G)$.
Chandran and Subramanian~\cite{spectralTreewidthLowerBd-chandran-subramanian-2003} showed that $\tw(G)+1\ge \frac{3\lambda_2(G)n}{4\Delta(G)+8\lambda_2(G)}$, where $\Delta(G)$ denotes the maximum degree of $G$.
Their bound has been very recently improved by Gima, Hanaka, Noro, Ono, and Otachi in \cite{SpectralTWLB-gima-hanaka-noro-ono-otachi-2024}, and subsequently in \cite{SpectralTWLB-gima-hanaka-noro-ono-otachi-2024-arXiv}, to $\tw(G)+1\ge \frac{\lambda_2(G)n}{\Delta(G)+\lambda_2(G)}$.
However, there have been no studies on upper-bounding $\tw(G)$ using the Laplacian spectrum.

\paragraph*{Treewidth of the line graph.}
As $\tw(G)$ and $\tw(L(G))$ are related via the inequality (see, for e.g., \cite{multicutsUnweighted-cualinescu-fernandes-reed-2003} for the left side and \cite{TreewidthLineGraph-harvey-wood-2018}, Proposition 2.3 for the right side)
\begin{equation}
    \Delta(G)(\tw(G)+1)\ge\tw(L(G))+1\ge \tw(G),
    \label{eq:intro-tw-L(G)-Delta(G)}
\end{equation}
one can obtain the following lower bound on $\cng(G)$ by using (\ref{eq:intro-congestion-tw(L(G))}) and the result of \cite{SpectralTWLB-gima-hanaka-noro-ono-otachi-2024-arXiv}:
$\cng(G)\ge \frac{\lambda_2(G)n}{\Delta(G)+\lambda_2(G)}$.
Another result that indirectly relates $\cng(G)$ with the algebraic connectivity is via Lemma 5.2 of \cite{SimulatingQuantumComp-markov-shi-2008}, which states that $\cng(G)\ge \frac{\alpha(G)\cdot n}4$ where $\alpha(G):=\min\limits_{S\subset V(G)}\frac{e_G(S,\overbar S)}{\min\{|S|,|\overbar S|\}}$ is the min-cut-ratio. 
Combined with the Cheeger inequality~\eqref{eq:cheegerLower}, this yields $\cng(G)\ge \frac{\lambda_2(G)n}{8}$.
Theorem~\ref{thm:MAINTHM} shows that $\cng(G)\ge \frac{2\lambda_2(G)n}{9}$, which is stronger than both of these bounds.

\subsection{Spectral clustering and Cheeger inequalities}
The spectral clustering algorithm considers the eigenvector $x$ of $\lambda_2(G)$ and partitions $V(G)$ into $S\sqcup \overbar{S}$, where $S=\{v\in V(G): x_v\ge 0\}$.
It is known to be a good approximation of finding $S^\ast:=\argmin_{S\subseteq V(G)} e_G(S,\overbar S)/\min\{|S|,|\overbar S|\}$, which is an NP-hard problem. \cite{LaplacianSurvey-mohar-1991, SpectralClustering-ng-jordan-weiss-2001}.
The quality of this approximation is quantified via the Cheeger inequality.

\begin{lemma}[Discrete Cheeger Inequalities~\cite{LaplacianSurvey-mohar-1991}]
\label{lem:CheegerIneq}
    For any graph $G$ and subset $S$ of vertices, we have
    \begin{equation}
        \label{eq:cheegerLower}
        \frac{\lambda_2(G)}{n}\le \frac{e_G(S,\overbar S)}{|S||\overbar S|} \le \frac{\lambda_n(G)}{n}.
    \end{equation}
    Moreover, if $(S,\overbar S)$ is the output of spectral clustering,
    \begin{equation}
        \label{eq:cheegerUpper}
        \frac{e_G(S,\overbar S)}{\min\{|S|,|\overbar S|\}}\le \sqrt{(2\Delta(G)-\lambda_2(G))\lambda_2(G)}.
    \end{equation}
\end{lemma}

Another useful measure of the sparsity of a cut is its \emph{conductance}.
For a subset $S\subseteq V(G)$, let $\Vol(S)=\sum_{v\in S}\deg_G(v)$ denote the volume of $S$.
Then, the conductance of $S$ is defined as $\phi_G(S):=\frac{e_G(S,\overbar S)}{\min\{\Vol(S),\Vol(\overbar S)\}}$.
Let $m$ denote the number of edges of $G$.
An exact analogue of Lemma~\ref{lem:CheegerIneq} can be demonstrated using the normalized Laplacian matrix $\mathcal L_G = I-D_G^{-1/2}A_GD_G^{-1/2}$.
If $0=\mu_1(G)\le \cdots\le \mu_n(G)\le 1$ are the eigenvalues of $\mathcal L_G$, then it is known that (see \cite{fourCheegerProofs-Chung-2007}, for example) 
for any $S\subseteq V(G)$, $\frac{\mu_2(G)}{2m}\le \frac{e_G(S,\overbar S)}{\Vol(S)\Vol(\overbar S)}\le \frac{\mu_n(G)}{2m}$,
and that by replacing the matrix $L_G$ by $\mathcal L_G$ in the spectral clustering procedure, we obtain a cut satisfying
$\phi_G(S)\le \sqrt{(2-\mu_2(G))\mu_2(G)}$.


\subsection{Minimum Linear Arrangement and Dynamic Programming}
The work of Ibrahim et al.~\cite{OptimalContractionTrees-ibrahim-et-al-2022} reduces the construction of high-quality contraction trees to a \emph{linear ordering} problem on the associated graph.
Given a fixed linear order $\sigma$ of the vertices, they show that an optimal binary contraction tree consistent with $\sigma$ (for several standard cost measures, including maximum intermediate tensor size and total operation count) can be found in polynomial time by a dynamic programming procedure, which also is the main frame for our Algorithm~\ref{def:Algo-SO-DP}.
The hard part is then shifted to finding a good order $\sigma$, for which they use a multilevel heuristic solver proposed in \cite{GraphMLA-safro-ron-brandt-2006} for the $p$-sum objective with $p=1$ (the NP-hard Minimum Linear Arrangement problem), implemented in their \texttt{LinearOrdering.jl} package~\cite{LinearOrdering.jl-code-ibrahim-2022}.
In practice, this combination of a Minimum Linear Arrangement heuristic with DP over the resulting order yields contraction trees that often outperform existing solvers on quantum-circuit benchmarks.
In our experiments we implement their method and use it as a strong baseline, and refer to it simply as the ``Ibrahim et al.'' method.

Our heuristic in Algorithm~\ref{def:Algo-SO-DP} bypasses the complexity of approximating a good minimum linear arrangement by simply using the spectral ordering as an easily available and efficient linear ordering, while giving effective contraction trees.
One can also think of spectral ordering as a relaxation to the minimum linear arrangement problem.

\subsection{Cotengra and Hyper-optimized Tensor network contraction}
Gray and Kourtis~\cite{HyperOptimizedTNContraction-gray-kourtis-2021} view a tensor network as a weighted hypergraph and construct contraction trees using a combination of recursive hypergraph partitioning (e.g. via KaHyPar~\cite{KaHyPar-schlag-henne-heuer-meyerhenke-sanders-schulz-2016}), greedy local heuristics, and problem-specific simplification and slicing strategies. 
They further introduce a \emph{hyper-optimization} layer that repeatedly samples contraction paths from a portfolio of heuristics while tuning their hyperparameters (for example, through Bayesian optimization), retaining the best path found. 
The open-source \texttt{cotengra} library~\cite{cotengra-lib} implements these ideas and exposes several practical optimizers, including a default automatic strategy (``Cotengra-Auto''), a generic hyper-optimizer (``Hyper-Opt''), and randomized greedy variants (``Hyper-Greedy'').
In our experiments, we compare our spectral order + dynamic programming (``SO+DP'') heuristic of Algorithm~\ref{def:Algo-SO-DP} directly against these baselines.

\section{Congestion bounds via Laplacian Eigenvalues}
\label{sec:congestion-bounds-via-laplacian}

\subsection{Proof of Theorem~\ref{thm:MAINTHM}}
In this subsection, we state and prove the main theorem of this work.
Recall that $\varepsilon(G)$ is the balance of the cut $S\sqcup\overbar S$ given by spectral clustering on $G$, i.e. $\varepsilon(G)=\min\left\{|S|/n, |\overbar S|/n\right\}$.

\begin{proof}[Proof of (\ref{eq:intro-thm-mainthm})]
Let $\ContractionTree$ be any contraction tree with an embedding $\pi:V(G)\to\ell(\ContractionTree)$.
Suppose $S$ is the furthest node from the root of $\ContractionTree$ with $|S|>2n/3$.
If $S_1$ and $S_2$ are the children of $S$, then as $|S_1|+|S_2|>2n/3$, one of them, say $S_1$, satisfies $|S_1|\ge n/3$.
As the distance from $S_1$ to the root of $\ContractionTree$ is larger than that from $S$, we also have $|S_1|\le 2n/3$.

The rank of the tensor corresponding to $S_1$ is $e_G(S_1,\overbar{S_1})$.
As $n/3\le |S_1|\le 2n/3$, we have 
\begin{equation}|S_1|(n-|S_1|)\ge \frac{n}3\cdot \frac{2n}3=\frac{2n^2}9.\end{equation}
Thus,
\begin{equation}
\begin{aligned}
\cng_{\pi,\ContractionTree}(G)\ge e_G(S_1,\overbar{S_1}) \stackrel{\text{(\ref{eq:cheegerLower})}}{\ge}  \lambda_2(G)\cdot \frac{|S_1|(n-|S_1|)}{n} \ge \lambda_2(G)\cdot \frac{2n}9,
\end{aligned}
\end{equation}
completing the proof of the left side of (\ref{eq:intro-thm-mainthm}). \hfill{$\blacksquare$}

To prove the right side of (\ref{eq:intro-thm-mainthm}), note that for any pair $(\pi, \ContractionTree)$, $\cng_{\pi,\ContractionTree}(G)=e_G(S,\overbar S)$ for some $S\subseteq V(G)$.
Thus, the desired inequality follows:
\begin{equation}
    \frac{\cng_{\pi,\ContractionTree}(G)}n\stackrel{\text{(\ref{eq:cheegerLower})}}{\le} \lambda_n(G)\cdot \frac{|S|(n-|S|)}{n^2} \le \frac{\lambda_n(G)}4.
\end{equation}
\end{proof}

In order to prove (\ref{eq:intro-thm-mainthm-3}) and (\ref{eq:intro-thm-mainthm-2}), we need to find contraction orders with low congestion.

\begin{proof}[Proof of (\ref{eq:intro-thm-mainthm-3})]
Consider a simple contraction procedure that equipartitions $G$ into parts $S_1\sqcup S_2\sqcup S_3$, and recursively equipartitions each subset into equal halves.
We will call this procedure $(\pi_{\mathrm{bal3}}, \ContractionTree_{\mathrm{bal3}})$.
Let $\pi_{\mathrm{bal3}}$ embed $S_1$ and $S_2\sqcup S_3$ into the first level of $\ContractionTree_{\mathrm{bal3}}$, then $S_2$ and $S_3$ into the third level of $\ContractionTree_{\mathrm{bal3}}$ below $S_2\sqcup S_3$, and continue recursively halving each subset, and embedding each half into the children of the original subset.
For this embedding, for each $S\in\{S_1,S_2,S_3\}$ we have $e_G(S,\overbar S)\le \lambda_n(G)\cdot \frac{|S||\overbar S|}n\le 2n\lambda_n(G)/9$, and for every other $S\in V(\ContractionTree)$ we have $e_G(S,\overbar S)\le \lambda_n(G)\cdot \frac{|S||\overbar S|}{n} < 2n\lambda_n(G)/9$.
Thus, this embedding satisfies $\cng_{\pi_{\mathrm{bal3}},\ContractionTree_{\mathrm{bal3}}}(G)\le 2n\lambda_n(G)/9$, as desired.
\end{proof}

\begin{proof}[Proof of (\ref{eq:intro-thm-mainthm-2})]
Let $\ContractionTree$ be the contraction ordering obtained by first using spectral clustering to divide $G$ into two parts $(S, \overbar S)$, followed by recursively equipartitioning each part in approximately equal halves.
By (\ref{eq:cheegerUpper}), the rank of $S$ is at most $\sqrt{(2\Delta(G)-\lambda_2(G))\lambda_2(G)}\cdot\min\{|S|,|\overbar S|\}$.
By the definition of $\varepsilon(G)$, we get
\begin{equation}
\label{eq:proof-upper-spectral}
    e_G(S,\overbar S)\le \sqrt{(2\Delta(G)-\lambda_2(G))\lambda_2(G)}\cdot \varepsilon(G)\cdot n.
\end{equation}
On the other hand, if $B\in \ContractionTree$ is a descendant of $S$ or $\overbar S$ with distance at least two from the root node, it can be seen that $|B|\le \left\lceil (1-\varepsilon(G))n/2\right\rceil\le \frac{(1-\varepsilon(G))n}2+\frac12$.
By (\ref{eq:cheegerLower}), this implies
\begin{equation}
\label{eq:proof-upper-hybrid}
    \begin{aligned}
    e_G(B,\overbar B)&\le \frac{\lambda_n(G)}n\cdot \left(\frac{(1-\varepsilon(G))n}2+\frac12\right)\left(n-\frac{(1-\varepsilon(G))n}2-\frac12\right)\\
    &<{\lambda_n(G)}\cdot\left(\frac{n(1-\varepsilon(G)^2)+1}4\right).
    \end{aligned}
\end{equation}

Combining (\ref{eq:proof-upper-spectral}) and (\ref{eq:proof-upper-hybrid}) gives us the desired bound in (\ref{eq:intro-thm-mainthm-2}).
\end{proof}

\subsection{Normalized Laplacian}
\label{sec:extension-normalizedL}
We now sketch an analogous version of Theorem~\ref{thm:MAINTHM} in terms of the eigenvalues of the normalized Laplacian $\mathcal L_G = I - D_G^{-1/2}A_GD_G^{-1/2}$.
Recall that the eigenvalues of $\mathcal L_G$ satisfy $0=\mu_1(G)\le \mu_2(G)\le \cdots \le \mu_n(G)\le 1$.
Then, the following holds.
\begin{theorem}
\label{thm:normalizedMAINTHM}
    For any $G$ and embedding $\pi:V(G)\to\ell(\ContractionTree)$,
    \begin{equation}
    \label{eq:normalized-mainthm-1}
        \frac{2\mu_2(G)}9\le \frac{\cng_{\pi,\ContractionTree}(G)}{2m}\le \frac{\mu_n(G)}4.
    \end{equation}
    Moreover, there are embeddings $\pi_i:V(G)\to\ell(\ContractionTree_i)$ such that
    \begin{align}
        \frac{\cng_{\pi_1, \ContractionTree_1}(G)}{2m}&\le \frac{2\mu_n(G)}{9}, \text{ and } \label{eq:normalized-mainthm-3}\\
        \frac{\cng_{\pi_2, \ContractionTree_2}(G)}{2m}&\le \max\left\{\varepsilon'(G) \sqrt{(2-\mu_2(G))\mu_2(G)}\, ,\,  \left(\frac{1-\varepsilon'(G)^2+1/(2m)}4\right) \mu_n(G)\right\}. \label{eq:normalized-mainthm-2}
    \end{align}
    Here $\varepsilon'(G)$ is the balance of the normalized spectral clustering algorithm.
\end{theorem}

We omit the proof of Theorem~\ref{thm:normalizedMAINTHM}, as it is exactly analogous to that of Theorem~\ref{thm:MAINTHM}, with $n$, $\varepsilon(G)$, $|S|$ and $|\overbar S|$ replaced by $2m$, $\varepsilon'(G)$, $\Vol(S)$ and $\Vol(\overbar S)$, respectively.
It is also worth noting that when $G$ is a regular graph, say of degree $d$, the bounds in Theorem~\ref{thm:normalizedMAINTHM} exactly translate to those in Theorem~\ref{thm:MAINTHM}, as in that case, $\mathcal L = L/d$, implying $\varepsilon(G)=\varepsilon'(G)$, $\lambda_i(G)=d\mu_i(G)$ and $2m=dn$.

\section{Congestions of specific graph families}
\label{sec:theoretical-congestion-bounds}
We focus on congestion for several graph families that frequently appear in physics and computer science.
First, we consider graphs with regular geometries such as hypercubes and lattices.
Next, we consider random regular graphs and Erd\H os-R\'enyi graphs.
These are fundamental random graph models, and moreover, have good eigenvalue concentration properties, making them a good candidate for congestion bounds.
In order to describe the asymptotic behavior of our results, we use standard asymptotic notations of $O$, $o$, $\Omega$, and $\omega$.

\subsection{Hypercubes}
For $d\ge 2$, the $d$-dimensional hypercube $Q_d$ has vertex set $\{0,1\}^d$ and edges between vertices that differ in exactly one coordinate.
It is well-known that the Laplacian eigenvalues of $Q_d$ are $0, 2, 4, \ldots, 2d$ with multiplicities $\binom d0, \binom d1,\ldots, \binom dd$ respectively~\cite{HypercubeSpectrum-chen-zhao-han-2021}.
So, Theorem~\ref{thm:MAINTHM} yields the universal lower bound $\cng(Q_d)\ge \frac{2\lambda_2(Q_d)}9\cdot 2^d=\frac49\cdot 2^d$ and the upper bound of $\left(\frac 38+\frac1{2^{d-1}}\right)dn$.
However, one can determine the congestion of the hypercube exactly as follows.

\begin{prop}[Congestion of the hypercube]
\label{prop:hypercube-exact-cng}
For every $d\ge 2$, $\cng(Q_d)=2^{d-1}$.
\end{prop}

\begin{proof}
\emph{Upper bound.}
Consider the contraction tree obtained by repeatedly cutting along coordinates, i.e., every internal bag is a subcube obtained by fixing some set of coordinates.

If a bag $B$ fixes exactly $r$ coordinates, then $|B|=2^{d-r}$ and every fixed coordinate contributes exactly $2^{d-r}$ cut edges leaving $B$, hence $|\partial B| = r\cdot 2^{d-r}$.
Since $\max\{r/2^r : r\ge 1\}$ is $1/2$ (attained at $r=1,2$), we have $|\partial B|\le 2^{d-1}$ for every bag, and therefore $\cng(Q_d)\le 2^{d-1}$.

\emph{Lower bound.}
Let $(\pi,\ContractionTree)$ be any contraction order on $Q_d$.
As in the proof of (\ref{eq:intro-thm-mainthm}), $\ContractionTree$ contains a bag $B$ with $2^d/3 \le |B|\le 2^{d+1}/3$.
By Harper's edge-isoperimetric inequality for the hypercube~\cite{IsoperimetricHypercube-Harper-1966}, for $|B|\le 2^{d-1}$ one has
\[
|\partial B|\ \ge\ |B|\log_2\left(\frac{2^d}{|B|}\right),
\]
and the same bound for $|B|\ge 2^{d-1}$ follows by applying it to the complement.
Over the interval $|B|\in[2^d/3,2^{d+1}/3]$, the minimum of the right-hand side occurs at $|B|=2^{d-1}$, giving $|\partial B|\ge 2^{d-1}$.
Hence $\cng(Q_d)\ge 2^{d-1}$.
\end{proof}



\subsection{Non-periodic and Periodic Lattices}
For two graphs $H_1$ and $H_2$, the Cartesian product $H_1\square H_2$ has vertex set $\{(u, v): u\in H_1, v\in H_2\}$, and $(u_1, v_1)$ and $(u_2, v_2)$ are adjacent if $u_1=u_2$ and $v_1v_2\in E(H_2)$, or $v_1=v_2$ and $u_1u_2\in E(H_1)$.
The Laplacian spectrum of $H_1\square H_2$ was studied in \cite{AlgebraicConnectivity-fiedler-1973}, and is known to be given by $\{\lambda_i(H_1)+\lambda_j(H_2): 1\le i\le |V(H_1)|, 1\le j\le |V(H_2)|\}$.

Suppose $P_k$ and $C_k$ denote the path and cycle graphs with $k$ vertices.
Then, the $m\times n$ non-periodic grid is given by the product $P_m\square P_n$, and the periodic grid is given by $C_m\square C_n$.
We observe that Theorem~\ref{thm:MAINTHM}, along with the fact that the Laplacian eigenvalues $P_k$ are $\{4\sin^2(q\pi/2k): 0\le q\le k-1\}$ and that of $C_k$ are $\{4\sin^2(q\pi/k): 0\le q\le k-1\}$, gives us spectral lower/upper bounds for these lattices.
Unfortunately, the spectral bounds thus obtained are weak.
In the following proposition, we explicitly determine the congestion of each of these lattices.

\begin{prop}[Congestion of lattices]
\label{prop:lattice-exact-cng}
For $m, n\ge 2$, we have
\begin{equation}
\begin{aligned}
    &\cng(P_m\square P_n)=\min\{m,n\}, \text{ and }\\&\cng(C_m\square C_n)\in\left\{2\min\{m,n\}-i: i\in\{0,1,2\}\right\}.
\end{aligned}
\end{equation}
Moreover, when $|m-n|\ge 2$, $\cng(C_m\square C_n)=2\min\{m,n\}$.
\end{prop}

\begin{proof}
    Without loss of generality, suppose $m\ge n$.
    
    \emph{Upper bounds.}
    For $P_m\square P_n$, consider the ``peeling'' contraction tree that merges columns from left to right:
    at some stage the bag $B_k$ consists of the first $k$ columns, i.e. a $k\times n$ subgrid.
    Then the only edges leaving $B_k$ are the $n$ horizontal edges between column $k$ and column $k+1$, so $|\partial B_k|=n$ for all $1\le k\le m-1$.
    Thus this embedding has congestion $n$, proving $\cng(P_m\square P_n)\le n$. \hfill{$\blacksquare$}

    For $C_m\square C_n$, we use the analogous peeling order along the cycle direction:
    let $B_k$ be $k$ consecutive columns of the torus.
    Now there are \emph{two} interfaces to the complement (left and right around the cycle), each contributing $n$ edges, hence $|\partial B_k|=2n$ for all $1\le k\le m-1$.
    Therefore $\cng(C_m\square C_n)\le 2n$. \hfill{$\blacksquare$}

    \emph{Lower bounds.}
    By Eq. (\ref{eq:intro-congestion-tw(L(G))}), $\cng(G)=\tw(L(G))+1$, and Eq. (\ref{eq:intro-tw-L(G)-Delta(G)}) implies $\tw(L(G))+1\ge \tw(G)$.
    Thus, $\cng(G)\ge \tw(G)$.
    It is classical that $\tw(P_m\square P_n)=n$~\cite{Ratcatcher-seymour-1994}.
    For the periodic lattices, it is known that (\cite{TreewidthPeriodicLattice-aidun-dean-morrison-yu-yuan-2020}, Theorem 1.2)
    \[
    \tw(C_m\square C_n)=\begin{cases}
        2\min\{m,n\} & |m-n|\ge 2,\\
        2\min\{m,n\} \text{ or } 2\min\{m,n\}-1 & |m-n|=1,\\
        2n-1 \text{ or } 2n-2 & m=n.
    \end{cases}
    \]
    This finishes the proof of Proposition~\ref{prop:lattice-exact-cng}.
    When $m\ge n+2$, the congestion is exactly $2n$.
    \hfill{$\blacksquare$}
    
\end{proof}

\subsection{Random Regular Graphs}

Let $\mathcal G(n,d)$ denote the uniform random $d$-regular graph on $n$ vertices (with $d$ fixed and $n$ large), and let $\cng(G)$ denote the minimum congestion over all contraction-tree embeddings considered in Theorem~\ref{thm:MAINTHM}.
When $d$ is fixed and $n$ is large, the eigenvalues of the adjacency matrix of a randomly selected graph from $\mathcal G(n,d)$ were given by McKay~\cite{expectedEigenvalueDistribution-mckay-1981}.
It is known that the largest adjacency eigenvalue is $d$, and all other eigenvalues are bounded by approximately $2\sqrt{d-1}$ in magnitude.
As $L_G+A_G=dI$ by definition, the eigenvalues of $A_G$ are exactly $d-\lambda_i(G)$, and the eigenvectors of $L_G$ and $A_G$ are the same.
Therefore, these concentration results on the adjacency spectrum translate to the Laplacian spectrum in the following form.
\begin{lemma}[Friedman, \cite{proofOfAlonsSecondEigenvalueConjecture-Friedman-2003}]
    \label{lem:friedman-RRG-eigs}
    Let $\epsilon>0$ be any constant.
    Then, with probability $1-o(1)$ (as $n\to\infty$), the Laplacian spectrum of $\mathcal G(n,d)$ satisfies
    \begin{equation}
    \label{eq:experiments-RRG-FriedmanBound}
        d-2\sqrt{d-1}-\epsilon \le \lambda_2(G)\le \cdots \le \lambda_n(G)\le d+2\sqrt{d-1}+\epsilon.
    \end{equation}
\end{lemma}

However, for any fixed $d$, random regular graphs are strong expanders and do not possess a meaningful low-conductance 2-way spectral cut.
In particular, plugging the Friedman eigenvalue bounds into Theorem~\ref{thm:MAINTHM} yields bounds on $\cng(G)$ of the form $\cng(G) = nd \pm n\sqrt{d-1} + O(n)$ arising from the Laplacian spectrum, and hence not tight for this random model.
Instead, we will show Theorem~\ref{thm:RRG-ERG-Congestion} for random regular graphs by reusing our contraction procedure $\pi_{\mathrm{bal3}}, \ContractionTree_{\mathrm{bal3}}$.
Recall that it partitions $V$ into three parts $S_1,S_2,S_3$ of sizes $\lfloor n/3\rfloor,\lceil n/3\rceil$ and then recursively equipartitions each $S_i$ into halves until singletons, yielding a binary contraction tree.

\begin{lemma}[Cut concentration for a fixed tree in $\mathcal G(n,d)$]
\label{lem:rrg-tree-cut-conc}
Fix any deterministic binary partition tree $\ContractionTree$ on $\{1,\ldots, n\}$ with at most $n-1$ internal nodes, and let $\mathcal B(\ContractionTree)$ be the family of vertex sets appearing as bags in $\ContractionTree$.
Let $G\sim\mathcal G(n,d)$ with $d$ fixed.
Then there exists an absolute constant $C>0$ such that with probability $1-o(1)$,
\begin{equation}
    \label{eq:rrg-tree-cut-conc}
    \max_{B\in\mathcal B(\ContractionTree)} \Bigl| |\partial B| - \frac{d\,|B|(n-|B|)}{n-1}\Bigr|\le C\sqrt{dn\log n}.
\end{equation}
\end{lemma}

\begin{proof}[Proof sketch]
We work in the configuration model $\mathcal G^*(n,d)$ and then condition on simplicity.\footnote{$\mathcal G^{*}(n,d)$ first generates a random multigraph with degree sequence $\mathbf d=(d,\dots ,d)$ by assigning $d$ ``stubs'' (half-edges) to each vertex and pairing them uniformly at random. Then we restrict the sample space to graphs with no self-loops and no multiple edges between any two vertices.}
For a fixed set $B$, the random variable $|\partial B|$ is a function of the random perfect matching on the $dn$ half-edges.
Exposing the matching pair-by-pair produces a Doob martingale in which changing one pairing affects $|\partial B|$ by at most $1$.
Then by Azuma--Hoeffding bound there is an absolute constant $c$ for which,
\[
\Pr\left(\Bigl||\partial B|-\mathbb E(|\partial B|)\Bigr|\ge t\right) \le 2\exp\left(-c{t^2}/{dn}\right).
\]
Moreover $\mathbb E(|\partial B|) = \frac{d\,|B|(n-|B|)}{n-1}$ in the configuration model.
Plugging int $t=C\sqrt{dn\log n}$ for a suitable constant $C=2/\sqrt c$, we have 
\[
\Pr\left(\Bigl||\partial B|-\frac{d\,|B|(n-|B|)}{n-1}\Bigr|\ge C\sqrt{dn\log n}\right) \le 2\exp\left(-4\log n\right) = O(n^{-4}).
\]
A union bound over at most $|\mathcal B(\ContractionTree)|\le n$ bags yields (\ref{eq:rrg-tree-cut-conc}) for $\mathcal G^*(n,d)$ with probability at least $1-n^{-3}$.
Finally, the probability that $G^*(n,d)$ is simple is bounded away from $0$ for fixed $d$, so the same event holds with probability $1-o(1)$ in $\mathcal G(n,d)$ as well.
\end{proof}

\begin{proof}[Proof of Theorem~\ref{thm:RRG-ERG-Congestion} for $\mathcal G(n,d)$]
Every bag $B$ appearing in $\ContractionTree_{\mathrm{bal3}}$ has size $|B|\le n/3$ or $|B|\ge 2n/3$.
Hence the deterministic quantity $\frac{d\,|B|(n-|B|)}{n-1}$ is maximized when $|B|\in\{\lfloor n/3\rfloor,\lceil n/3\rceil\}$, and the maximum equals $\frac{d}{n-1}\cdot \frac{2}{9}n^2 = \frac{2}{9}dn\cdot \frac{n}{n-1} = \frac{2}{9}dn+o(n)$.
Applying Lemma~\ref{lem:rrg-tree-cut-conc} with $\ContractionTree=\ContractionTree_{\mathrm{bal3}}$ gives us
$\cng_{\pi_{\mathrm{bal3}}, \ContractionTree_{\mathrm{bal3}}}(G)\le \frac{2}{9}dn + C\sqrt{dn\log n}$.

Moreover, for any contraction tree $\ContractionTree$, our argument in the proof of Theorem~\ref{thm:MAINTHM} showed that there is always a bag $B$ of size $n/3\le |B|\le 2n/3$, implying
$\cng_{\pi,\ContractionTree}(G)\ge \frac 29 dn - C'\sqrt{dn\log n}$,
finishing the proof of Theorem~\ref{thm:RRG-ERG-Congestion} for $\mathcal G(n,d)$.
\end{proof}

Theorem~\ref{thm:RRG-ERG-Congestion} matches the natural heuristic for random regular graphs: the leading term $\frac{2}{9}dn$ arises from the largest cut induced by the initial $\frac13$-$\frac23$ split, while the recursive equipartition cuts are smaller.

\subsection{Erd\H os-R\'enyi Random Graphs}

Let $G\sim \mathcal G_{n,p}$ be a graph sampled from the Erd\H os-R\'enyi random graph model with $n$ vertices and probability of adjacency $p$.
The Laplacian spectrum of such a graph $G$ is well-studied in the theory of random matrices, especially in the sparse regime $p=o(\log n/n)$ and dense regime $p=\omega(\log n/n)$.
In the sparse regime, $\mathcal G_{n,p}$ is almost surely disconnected.
On the other hand, studying the $\mathcal G_{n,p}$ spectrum for $p=\Theta(\log n/n)$ is a challenging task, see \cite{GnpSpectrum-kolokolnikov-osting-vonbrecht-2014} for more details.
Thus, here we consider the congestion of $\mathcal G_{n,p}$ in the dense regime  $p=\omega(\log n/n)$.

In the dense regime, the behavior of $\mathcal G_{n,p}$ follows that of the random regular graphs $\mathcal G(n,pn)$ with high probability, so we expect these random graph families to exhibit similar behavior.
Indeed, the behavior of the spectrum of $\mathcal G_{n,p}$ is known.
\begin{lemma}[Chung and Radcliffe~\cite{GnpSpectrum-chung-radcliffe-2011}, Ding and Tiang~\cite{GnpSpectrum-ding-tiang-2010}]
    Let ${pn}/{\log n}\to\infty$ as $n\to\infty$, and $G\sim\mathcal G(n,p)$. Then, with probability $1-o(1)$,
    \begin{equation}
        \label{eq:experiments-Gnp-Lambda_2-Lambda_n}
        |\lambda_i(G)-pn|\le O\left(\sqrt{pn\log n}\right)\text{ and } |\mu_i(G)-1|\le O\left(\sqrt{\frac{\log n}{np}}\right).
    \end{equation}
\end{lemma}
Combining (\ref{eq:experiments-Gnp-Lambda_2-Lambda_n}) and (\ref{eq:intro-thm-mainthm}) leads us to the following estimation of the congestion with probability $1-o(1)$: $\cng(G) = \frac29 pn^2 + O(\sqrt{pn^3\log n})$.
However, a completely parallel proof to the random regular graph case can be used to show an upper bound on $\cng_{\pi_{\mathrm{bal3}}, \ContractionTree_{\mathrm{bal3}}}$.

\begin{lemma}[Cut concentration for a fixed tree in $\mathcal G_{n,p}$]
\label{lem:gnp-tree-cut-conc}
Fix any deterministic binary partition tree $\ContractionTree$ on $\{1,\ldots, n\}$ with at most $n-1$ internal nodes, and let $\mathcal B(\ContractionTree)$ be the family of vertex sets appearing as bags in $\ContractionTree$.
Let $G\sim\mathcal G_{n,p}$ and assume $p=\omega(\log n/n)$.
Then there exists an absolute constant $C>0$ such that with probability $1-o(1)$,
\begin{equation}
\label{eq:gnp-tree-cut-conc}
\max_{B\in\mathcal B(\ContractionTree)}
\Bigl|\,|\partial B| - p\,|B|(n-|B|)\Bigr|
\;\le\;
C\sqrt{p n^2\log n}.
\end{equation}
\end{lemma}

\begin{proof}[Proof sketch]
Fix a set $B\subseteq\{1,\ldots, n\}$.
Then $|\partial B|$ is a sum of $|B|(n-|B|)$ independent $\mathrm{Bernoulli}(p)$ random variables,
so $\mathbb E|\partial B|=p|B|(n-|B|)$.
By the Chernoff bound, 
\[
\Pr\Bigl(\bigl||\partial B|-\mathbb E|\partial B|\bigr|\ge t\Bigr)\le 2\exp\left(-\frac{ct^2}{p|B|(n-|B|)+t}\right),
\]
for any $t\ge 0$. 
Taking $t=C\sqrt{pn^2\log n}$ for a suitable constant $C$, and using $|B|(n-|B|)\le n^2/4$, we obtain
\[
\Pr\Bigl(\bigl||\partial B|-p|B|(n-|B|)\bigr|\ge C\sqrt{pn^2\log n}\Bigr)\le O(n^{-4}).
\]
A union bound over at most $|\mathcal B(\ContractionTree)|\le n$ bags yields \eqref{eq:gnp-tree-cut-conc} with probability at least $1-n^{-3}$, which is $1-o(1)$.
\end{proof}
\begin{proof}[Proof of Theorem~\ref{thm:RRG-ERG-Congestion} for $\mathcal G_{n,p}$]
Every bag $B$ appearing in $\ContractionTree_{\mathrm{bal3}}$ has size $|B|\le n/3$ or $|B|\ge 2n/3$, implying $\cng_{\pi_{\mathrm{bal3}}, \ContractionTree_{\mathrm{bal3}}}(G) \le \frac{2}{9}pn^2 + C\sqrt{pn^2\log n}$.
Moreover, for any contraction tree $\ContractionTree$ and embedding $\pi$, there exists a bag $B\in V(\ContractionTree)$ with $n/3\le |B|\le 2n/3$, implying that with probability $1-o(1)$,
\[
|\partial B|\ge\frac{2}{9}pn^2 - O(pn) - C\sqrt{pn^2\log n} = \frac{2}{9}pn^2 - C'\sqrt{pn^2\log n},
\]
where we used $p=\omega(\log n/n)$ to absorb the $O(pn)$ term into the $\sqrt{pn^2\log n}$ term.
Since $\cng_{\pi,\ContractionTree}(G)\ge |\partial B|$, this finishes proof of Theorem~\ref{thm:RRG-ERG-Congestion} for $\mathcal G_{n,p}$.
\end{proof}

\section{Numerical Experiments}
\label{sec:experiments}
In this section, we compare the performance of Algorithm~\ref{def:Algo-SO-DP}, based on spectral ordering and dynamic programming, with the Hyper-Greedy, Cotengra-Auto, and Hyper-Optimized contraction orders of Gray and Kourtis~\cite{HyperOptimizedTNContraction-gray-kourtis-2021}, as well as the MLA-based approach of Ibrahim et al.~\cite{OptimalContractionTrees-ibrahim-et-al-2022}.
Our code is implemented in \texttt{python 3.12} and is available on GitHub~\cite{mukherjee-akiyama2025congestioncode}; all numerical computations were performed on an \texttt{Apple Silicon M4 Max} CPU with \texttt{64 GB} memory.
The experiments in this section are intended to illustrate that, while the general spectral upper bounds are often loose, they are universal and hold across all graph families, whereas the spectral lower bounds track the observed congestion qualitatively. A quantitative comparison of the tightness of these bounds is provided in Appendix~\ref{sec:appendix_Performance}.
Additional runtime comparisons are given in Appendix~\ref{sec:appendix_runtime}, where we note that Hyper-Opt~\cite{HyperOptimizedTNContraction-gray-kourtis-2021} consistently achieves the best contraction cost, but at the expense of an order-of-magnitude longer runtime.


\subsection{Non-periodic and Periodic Lattices}
\begin{figure}[h]
    \centering
        \includegraphics[width=\textwidth]{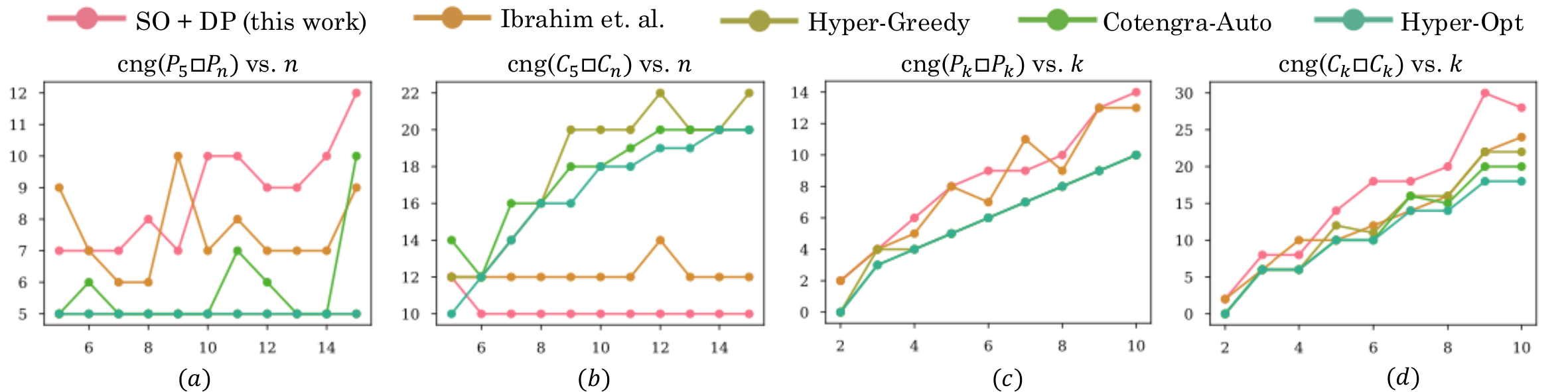}
    \caption{
    Comparison of the different bounds for: (a) the non-periodic lattice $P_m\square P_n$ and (b) periodic lattice $C_m\square C_n$ for $m=5$ and $5\le n \le 15$; (c) non-periodic square lattice and (d) periodic square lattice $P_k\square P_k$ for $5\le k \le 15$. In (a), Hyper-Greedy and Hyper-Opt exactly overlap. In (c), Cotengra-Auto and Hyper-Opt exactly overlap, and Hyper-Greedy also matches them for $k\ge 4$.}
    \label{fig:experiments-lattice}
\end{figure}
We compare the performances of different algorithms for $5\times n$ rectangular lattices ($5\le n\le 15$) in Fig.~\ref{fig:experiments-lattice} (a) and (b), and $k\times k$ square lattices ($5\le k\le 15$) in Fig.~\ref{fig:experiments-lattice} (c) and (d).
Our method outperforms the other techniques and finds the optimal embedding with $\cng(C_5\square C_5)=10$ for all $n\ge 6$ in the periodic rectangular lattice case.
In all other input graphs, Hyper-Opt gives the smallest congestion trees overall.
Especially for non-periodic square lattices, we find that all algorithms of \cite{HyperOptimizedTNContraction-gray-kourtis-2021} find the optimal contraction tree.



\subsection{Random Regular Graphs}
\begin{figure}[h]
    \centering
    \includegraphics[width=\textwidth]{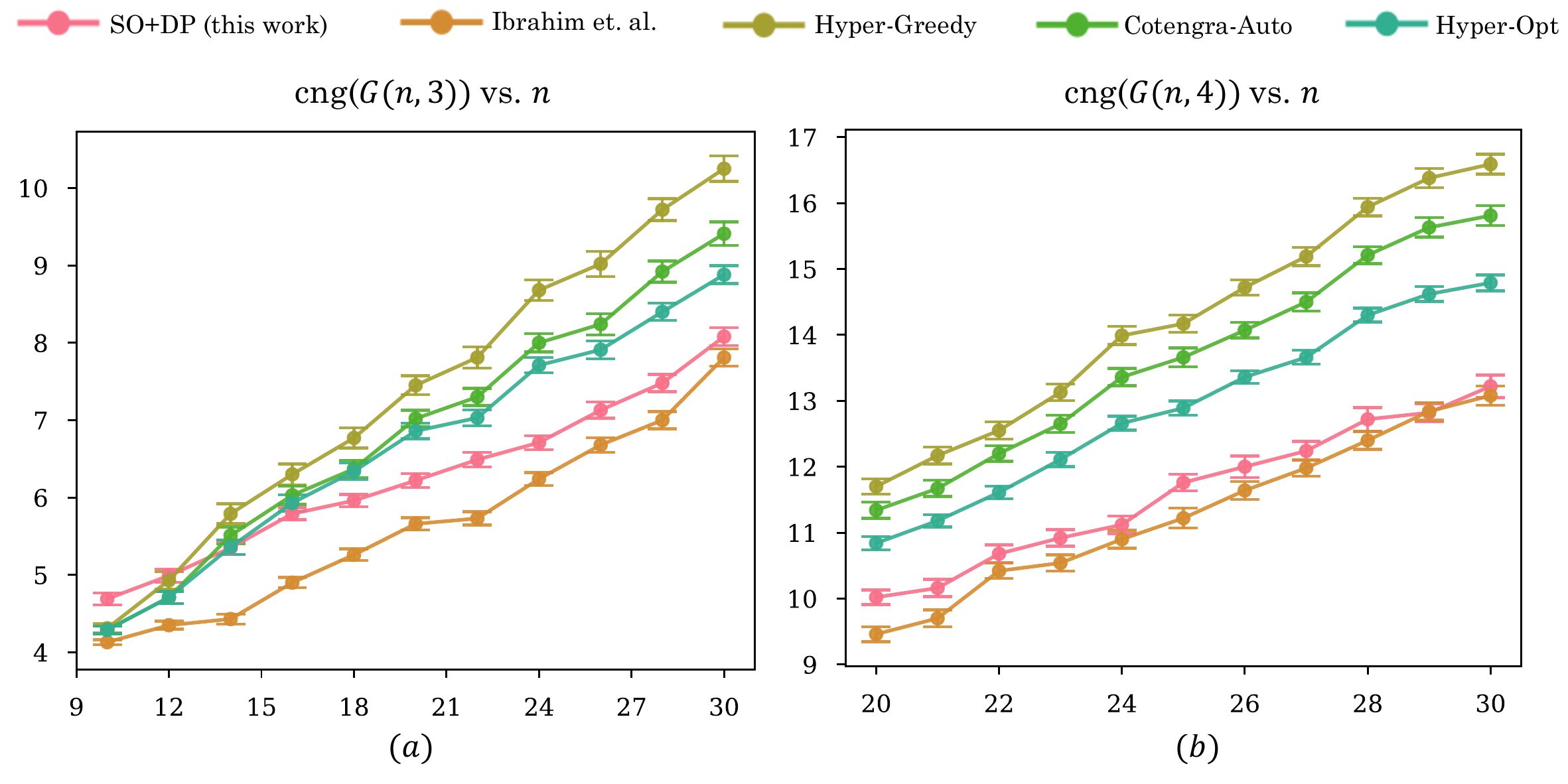}
    \caption{
    Congestion computed by different algorithms on $100$ instances of $\mathcal G(n,d)$, shown as means with $\pm1$ standard deviation error bars, for (a) $d=3$ and (b) $d=4$.
    }
    \label{fig:experiment-RRG}
\end{figure}

In Fig.~\ref{fig:experiment-RRG}, we analyze the performance of the different upper bounds on congestion for $d=3,4$ and increasing $n$.
While the MLA-based approach of \cite{OptimalContractionTrees-ibrahim-et-al-2022} performs well across the board, our proposed algorithm (Algorithm~\ref{def:Algo-SO-DP}) also performs comparably, and has reliably lower congestion than the contraction trees obtained by the Cotengra-based methods of \cite{HyperOptimizedTNContraction-gray-kourtis-2021} commonly used in practice.

\subsection{Erd\H os-R\'enyi Graphs}

\begin{figure}[h]
    \centering
    \includegraphics[width=\textwidth]{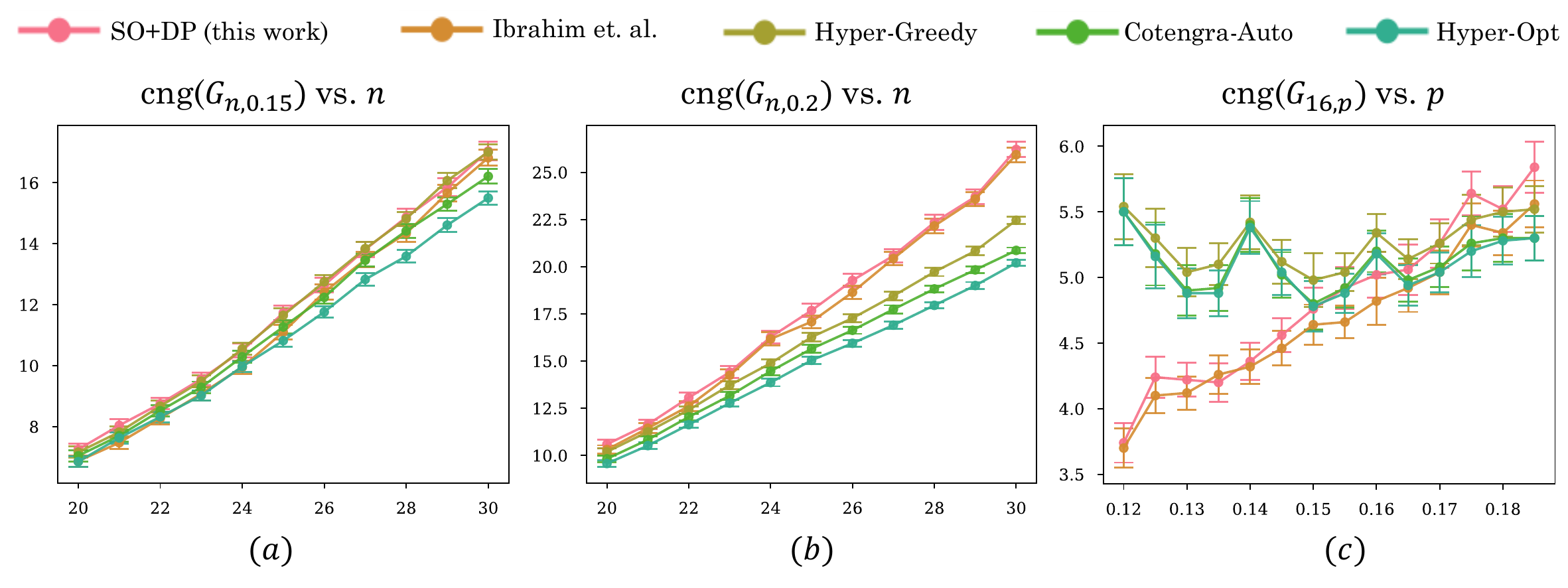}
    \caption{
    Congestion calculated by different algorithms for $100$ instances of $\mathcal G_{n,p}$, shown as means with $\pm1$ standard deviation error bars, for (a) $p=0.15$, (b) $p=0.2$, and (c) $n=16$, $0.12\le p\le 0.2$.}
    \label{fig:experiments-ERG}
\end{figure}

Fig.~\ref{fig:experiments-ERG} (a) and (b) compare different lower and upper bounds for $p=0.15, 0.2$ and increasing $n$.
It is noticeable that Hyper-Opt gives the best upper bound across the board when $\mathcal G_{n,p}$ is sufficiently dense.
However, we noticed in Fig.~\ref{fig:experiments-ERG} (c) that for relatively sparse $\mathcal G_{n,p}$'s, our proposed Algorithm~\ref{def:Algo-SO-DP} and the MLA-based approach~\cite{OptimalContractionTrees-ibrahim-et-al-2022} are comparable, and both better than the approaches of \cite{HyperOptimizedTNContraction-gray-kourtis-2021}.

Therefore, depending on the sparsity or density of the given graph, our proposed algorithm can play a complementary role to existing approaches.
Algorithm~\ref{def:Algo-SO-DP} seems to work well on relatively sparse input graphs.

\subsection{Quantum Circuits}
Finally, we turn our attention to quantum circuits.
Any quantum circuit can be represented as a tensor network, where the quantum gates and wires are described as tensors and edges between these nodes, respectively.
Every $k$-qubit gate in the circuit corresponds to a node of degree $2k$ in the tensor network, and by adding tensor nodes corresponding to the initial state and a target state, one can calculate the probability of measuring the target state by taking a tensor network contraction~\cite{SimulatingQuantumComp-markov-shi-2008}.
The memory cost of simulating quantum circuits using tensor networks is directly determined by the congestion of their graph representation.
Note that we are allowed to focus on quantum circuits without single-qubit gates because these gates do not contribute to the congestion.
In this section, we denote by $\mathrm{RQC}(q,d)$ a random quantum circuit with $q$ qubits of depth $d$ consisting of only $2$-qubit gates.

\begin{figure}[h]
    \centering
    \includegraphics[width=\linewidth]{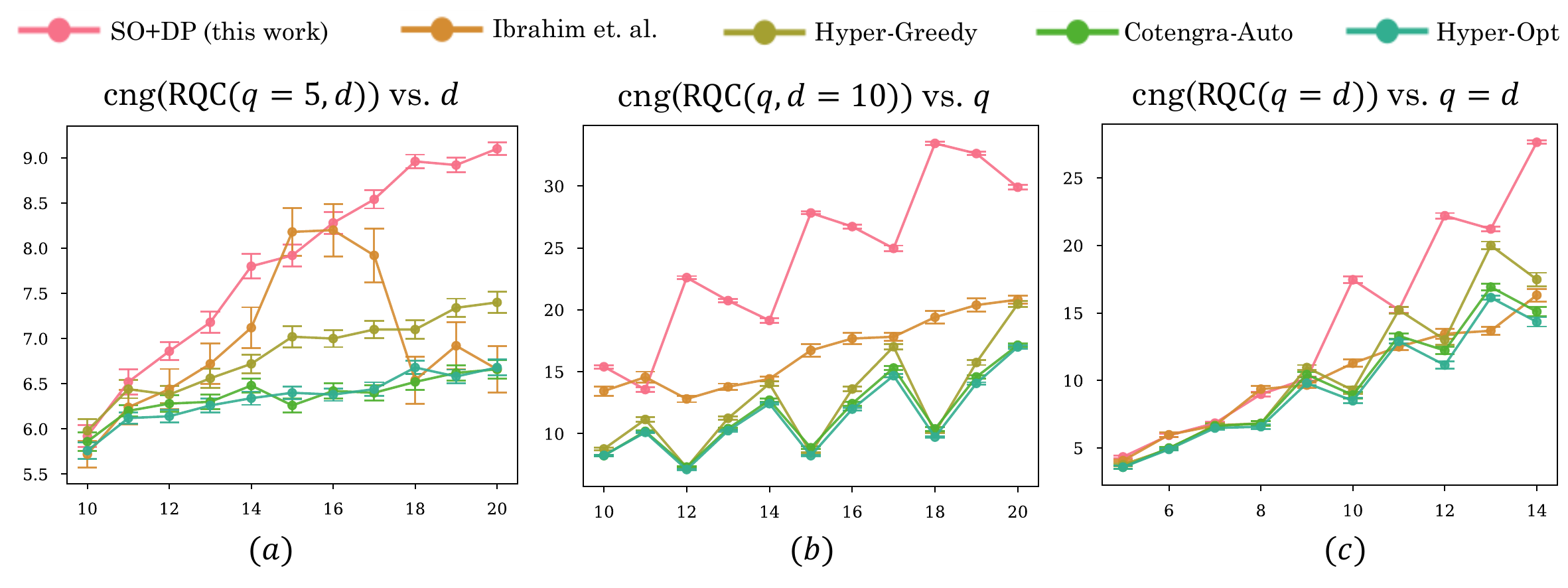}
    \caption{
    Congestion computed by different algorithms for $50$ instances of $\mathrm{RQC}(q,d)$ consisting of only $2$-qubit gates for:
    (a) $q=5$ with $10\le d\le 20$, 
    (b) $d=5$ with $10\le q\le 20$, and
    (c) $q=d$, shown as means with $\pm1$ standard deviation error bars.}
    \label{fig:experiments-randomQC}
\end{figure}

Fig.~\ref{fig:experiments-randomQC} shows the congestion calculated by different algorithms for various families of random quantum circuits consisting of only $2$-qubit gates.
The contraction procedures of \cite{HyperOptimizedTNContraction-gray-kourtis-2021} perform the best throughout the board for these graphs.

\section{Conclusion and Future Work}
\label{sec:conclusion-future-work}
In this work, we studied the complexity of tensor network contraction for arbitrary geometries by relating graph congestion to the unnormalized and normalized Laplacian spectra.
We proved general spectral lower and upper bounds on congestion, establishing for the first time a nontrivial upper bound, and exactly determined the congestion (or its asymptotics) for several fundamental graph families, including hypercubes, lattices, random regular graphs, and Erdős–Rényi graphs.
We also proposed a contraction procedure based on spectral ordering and dynamic programming, which is efficient and performs competitively on sparse graphs, complementing existing heuristic approaches.

In practical simulations, memory constraints are often alleviated through slicing, which trades peak memory for increased runtime. 
While slicing can arbitrarily reduce memory usage, it does not eliminate intrinsic bottlenecks in contraction orders and typically incurs a substantial increase in total arithmetic cost. 
So, we focus on the \emph{unsliced exact contraction} model, where congestion directly characterizes the minimum achievable peak memory. 
We think understanding the behavior of congestion for graphs is a first step towards analysis of space–time tradeoffs, which we leave for future work.

Finally, we would like to point out that the primary focus of this work is on the memory complexity of tensor network contraction, all our bounds on edge-congestion directly imply $1.5$-multiplicative factor bounds on vertex-congestion, and thus directly translate to time complexity.
Searching for other graph parameters representing congestion is a promising avenue for future work.



\bibliography{main.bbl}

\pagebreak

\appendix
\section{Performance of the spectral theoretical bounds}
\label{sec:appendix_Performance}

\begin{figure}[h]
    \centering
    \includegraphics[width=\textwidth]{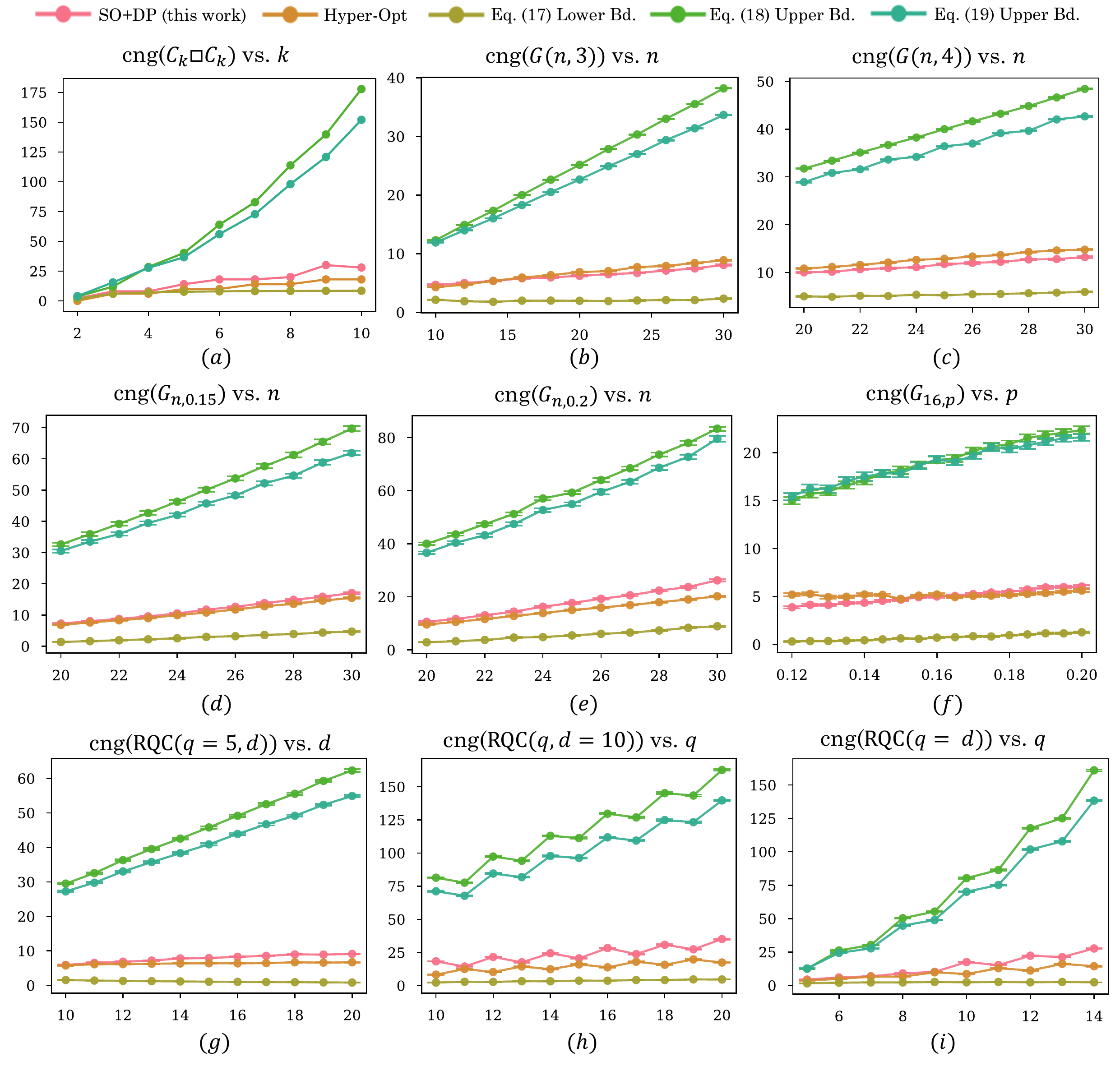}
    \caption{
    Congestion estimated by upper and lower bounds of Theorem~\ref{thm:normalizedMAINTHM}, in comparison with contraction trees produced by Algorithm~\ref{def:Algo-SO-DP} and Hyper-Opt.}
    \label{fig:appendix_soectral}
\end{figure}

In Fig.~\ref{fig:appendix_soectral}, we compare the performance of our general spectral bounds, which are: the lower bound of $4\mu_2(G)/9$ in Eq.~(\ref{eq:normalized-mainthm-1}), the generic upper bound of $\frac{2\mu_n(G)}{9}$ in Eq.~(\ref{eq:normalized-mainthm-3}), and its improved version via spectral balance considerations in Eq.~(\ref{eq:normalized-mainthm-3}) for the different graph families shown in Section~\ref{sec:experiments}.
It can be seen that in most cases, the upper bounds are very weak, whereas the lower bounds are closer to the expected congestion values.

\section{Runtime comparisons}
\label{sec:appendix_runtime}
In this appendix, we present the average runtime for a single input graph for Algorithm~\ref{def:Algo-SO-DP}, our implementation of Ibrahim et. al.~\cite{OptimalContractionTrees-ibrahim-et-al-2022}, Hyper-Greedy, Cotengra-Auto and Hyper-Opt~\cite{HyperOptimizedTNContraction-gray-kourtis-2021}.
We observe that while Hyper-Opt gives the best contraction trees in most situations, it also takes orders of magnitude longer compared to its competitors.
The measured average runtimes per instance for each of the random experiments from Section~\ref{sec:experiments} (in milliseconds) are shown in Table~\ref{table:app-runtimes}.

\begin{table}[]
\centering
\tiny
\begin{tabular}{cccccc}
\textbf{Graph family} & \textbf{SO + DP} & \textbf{Ibrahim et. al.} & \textbf{Hyper-Greedy} & \textbf{Cotengra-Auto} & \textbf{Hyper-Opt} \\ \hline \hline
$\mathcal G(n,3)$           & $11.883 \pm 0.28$ & $9.116 \pm 0.223$ & $1.185 \pm 0.026$ & $16.377 \pm 0.729$ & $2330 \pm 142.28$ \\
$\mathcal G(n,4)$           & $20.126 \pm 0.444$ & $16.1 \pm 0.214$ & $2.731 \pm 0.035$ & $13.707 \pm 0.198$ & $7990 \pm 452.895$ \\
$\mathcal G_{n,p=0.15}$     & $7.08 \pm 0.158$ & $2.915 \pm 0.069$ & $0.641 \pm 0.01$ & $8.924 \pm 0.345$ & $1630 \pm 101.498$ \\
$\mathcal G_{n,p=2}$        & $18.022 \pm 0.308$ & $14.853 \pm 0.229$ & $2.863 \pm 0.056$ & $21.744 \pm 0.76$ & $4080 \pm 359.607$ \\
$\mathcal G_{16,p}$         & $23.929 \pm 0.61$ & $18.422 \pm 0.274$ & $5.391 \pm 0.101$ & $30.512 \pm 0.99$ & $1050 \pm 766.665$ \\
$\mathrm{RQC}(5,d)$         & $15.497 \pm 0.351$ & $19.999 \pm 0.53$ & $1.905 \pm 0.048$ & $26.445 \pm 1.498$ & $1430 \pm 46.222$ \\
$\mathrm{RQC}(q,10)$        & $154.221 \pm 4.102$ & $209.236 \pm 5.269$ & $6.412 \pm 0.128$ & $105.332 \pm 9.01$ & $1180 \pm 745.931$ \\
$\mathrm{RQC}(q=d)$         & $80.161 \pm 4.137$ & $104.511 \pm 5.623$ & $3.637 \pm 0.133$ & $28.167 \pm 1.087$ & $3860 \pm 278.184$ \\ \hline
\end{tabular}
\caption{Average runtime (ms) with $\pm$1 sample standard deviation for the algorithms compared in Section~\ref{sec:experiments}.\label{table:app-runtimes}}
\end{table}

\end{document}